\documentclass[11pt]{article}
\usepackage{Blois,epsfig}
\usepackage{graphicx}

\def\be{\begin{equation}}
\def\ee{\end{equation}}
\def\bea{\begin{eqnarray}}
\def\eea{\end{eqnarray}}

\def\beq{\begin{equation}}
\def\eeq{\end{equation}}
\def\qb{$q\bar{q}~$}
\def\q2{$Q^2$}
\begin{document}

%\vspace*{2cm}
%\begin{center}
%\Large{\textbf{XIth International Conference on\\ Elastic and Diffractive 
%Scattering\\ Ch\^{a}teau de Blois, France, May 15 - 20, 2005}}
%\end{center}

\vspace*{2cm}
\title{MEASUREMENT OF LIGHT-CONE WAVE FUNCTIONS BY DIFFRACTIVE 
DISSOCIATION}

\author{DANIEL ASHERY }

\address{School of Physics and Astronomy, Sackler Faculty of Exact 
Science, \\
Tel Aviv University, Tel Aviv 69978, Israel}

\maketitle\abstracts{
The measurement of the pion light-cone wave function is revisited and
results for the Gegenbauer coefficients are presented. Mesurements of the
photon electromagnetic and hadronic wave functions are described and
results are presented.}

%\newpage

\section{The Pion Light Cone Wave Function}
\begin{figure}[h]\centering
\includegraphics[width=8cm]{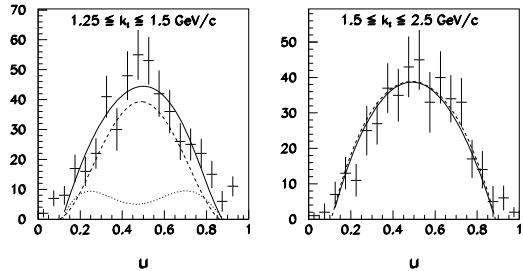}
\vglue -4.0cm
\caption{ The $u$ distribution of diffractive di-jets from the platinum 
target. The solid line is a fit to a
combination of the asymptotic and CZ wave functions. The dashed line shows
the contribution from the asymptotic function and the dotted line that of
the CZ function.}
\label{xdatadif}
\end{figure}

A differential measurement of the pion LCWF was performed by Fermilab
E791 collaboration \cite{791wf} by studying the diffractive dissociation
of 500 GeV/c pions, interacting with C and Pt targets, to two jets. If in
this process the quark momentum is transferred to the jet, measurement of
the jet momentum gives the quark (and antiquark) momentum. Thus:
$u_{measured} = \frac {p_{jet1}} {p_{jet1}+p_{jet2}}.$ It has been shown
\cite{fms} that the cross section for this
process is prportional to $\phi^2$. The resulting $u$ distributions are
shown in Fig. \ref{xdatadif} for  two windows of $k_t$:
$1.25 ~\rm{GeV/c} ~\leq ~k_t ~\leq ~1.5 ~\rm{GeV/c}$ and
$1.5 ~\rm{GeV/c} ~\leq  ~k_t ~\leq ~2.5 ~\rm{GeV/c}$. The results are
compared with linear combinations of simulations of squares of the
asymptotic \cite{bl} and CZ \cite{cz} distribution amplitudes.

\begin{figure}[h]\centering
\includegraphics[angle=0,width=8cm]{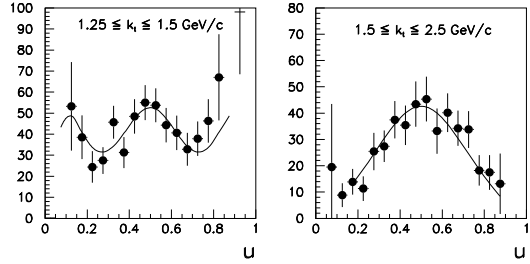}
\vglue -4.0cm
\caption{ The Acceptance-corrected $u$ distributions of diffractive
%di-jets obtained by applying correction to the E791 results \cite{791wf}.
di-jets obtained by applying correction to the E791 results$^1$.
The solid line is a fit to a combination of Gegenbauer polynomials, Eq.
1.}
\label{gegen}
\end{figure}

\begin{figure}[h]\centering
\includegraphics[angle=0,width=0.4\textwidth]{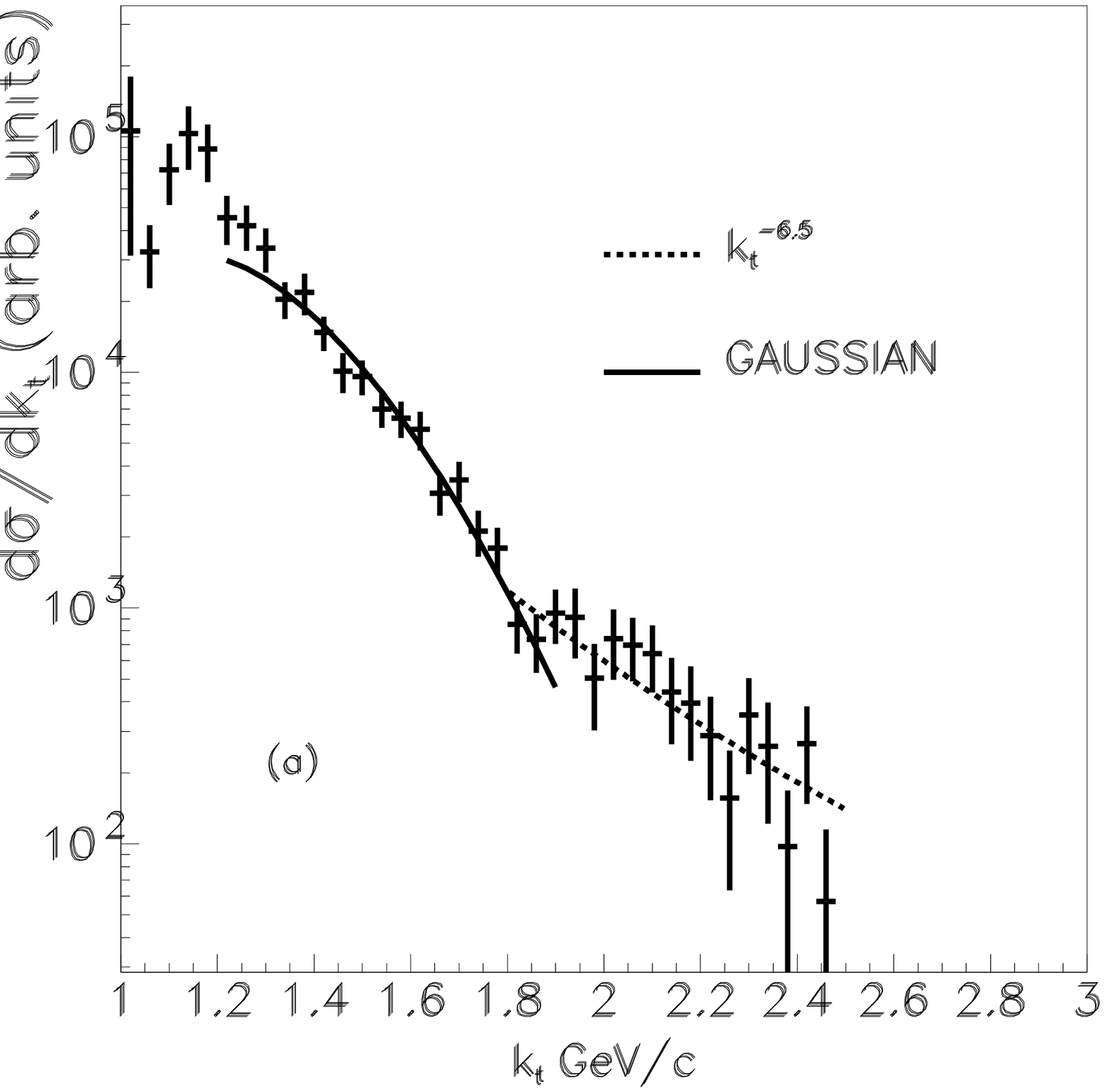}
  \hfill
\hspace{-1.0cm}
\includegraphics[angle=0,width=0.4\textwidth]{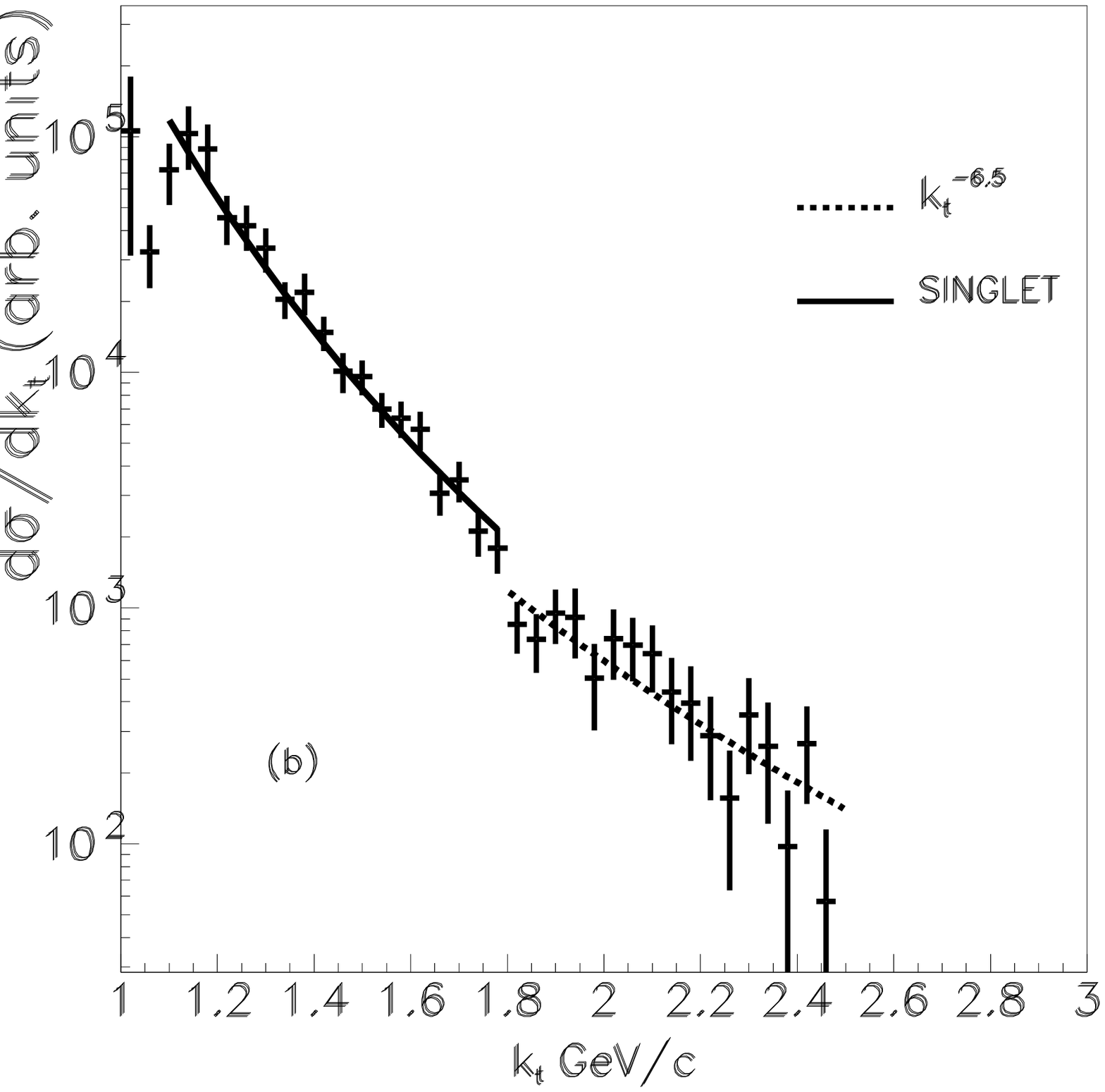}
%\caption{Comparison of the experimental $k_t$ distribution \cite{791wf}
\caption{Comparison of the experimental $k_t$ distribution$^1$
      with fits derived from:
%  (a) Gaussian LCWF \cite{krol} 
  (a) Gaussian LCWF$^6$ 
for low $k_t$ and a power law dependence:
      $\frac{d\sigma}{dk_t} \propto k_t^n$, as expected from
      perturbative calculations, for high $k_t$;
%  (b) Two-term Singlet model wave function \cite{dacp} 
  (b) Two-term Singlet model wave function$^7$
for low $k_t$ and
      a power law for high $k_t$. }
\label{figkt}
\end{figure}

The results for the higher $k_t$ window show that the asymptotic wave
function describes the data very well. Hence, for $k_t > $1.5 GeV/c, which
translates to $Q^2 \sim 10~{\rm (GeV/c)^2}$, the pQCD approach that
led to construction of the asymptotic wave function is reasonable.
The distribution in the lower window is consistent with a significant
contribution from the CZ wave function or may indicate contributions due
to other non-perturbative effects. A way to understand this better is
to use the experimental results to determine the coefficients of the
Gegenbauer polynomials in the expansion of $\phi_\pi(u,Q^2)$
\cite{bl}. One
cannot fit directly the results shown in Fig. \ref{xdatadif} as these
are distorted by the hadronization process and experimental
acceptance. It is also not practical to carry out simulations of the
distortions for a large variety of distribution amplitudes. We adopt here
a simpler approach, albeit somewhat less precise than that used by E791.
We use results of the Monte
Carlo simulations to correct the distortions and experimental acceptance
\cite{ppnp}. The results, Fig. \ref{gegen}, are fitted to the expression:
%\begin{eqnarray}
\begin{equation}
\frac{d\sigma}{du} \propto \phi^2_\pi(u,Q^2) = N\cdot u^2(1-u)^2  
\times\left(1.0 + a_2  C_2^{3/2}(2u-1)
  + a_4  C_4^{3/2}(2u-1)\right)^2 
\label{fit_geg}
%\end{eqnarray}
\end{equation}
where $N$ is a normalization constant and $C_n$ are the Gegenbauer
polynomials. The results of the fits are that for the
high $k_t$ region $a_2 = a_4 = 0$, confirming the conclusion
of the E791 authors that for this region $\phi_{Asy}^2$ describes the data
well. For the low $k_t$ region the coefficients are: $a_2 = 0.30 \pm 0.05,
~~a_4 = (0.5 \pm 0.1)\cdot 10^{-2}$.  The fact that $a_4 \neq 0$
indicates a distribution amplitude that is different from $\phi_{CZ}$
which contains only an $a_2$ term.
The $k_t$ dependence of diffractive di-jets is another observable that can
show the region where perturbative calculations describe the data. As
shown in \cite{fms} it is expected to be: ${d\sigma\over dk_t} ~\sim
~k_t^{-6}$. The results, shown in Fig. \ref{figkt}, are consistent with
this dependence only in the region above $k_t \sim$ 1.8 GeV/c, in
agreement with the conclusions from the $u$-distributions. For lower $k_t$
values, non-perturbative effects are expected to be significant.
Naturally, the transition between the two regions is not sharp. The region
of $1.0 \leq k_t \leq 1.8 ~GeV/c$ may be a transition region where we can
still apply pQCD techniques but must use LCWFs that better describe the
non-perturbative structure of the pion.
In Fig \ref{figkt}(a) it is compared with a Gaussian wave function
\cite{krol} and in Fig \ref{figkt}(b) to the a Singlet model wave function
\cite{dacp}. The observed sensitivity shows that the $k_t$ distribution is
useful in studying wave functions in this transition region.

\section{The Photon Light Cone Wave Function}

\subsection{Measurement of the electromagnetic component of real photon
LCWF }
\label{sec-dif}

\begin{figure}[h]
\begin{center}
\includegraphics[width=5.0cm]{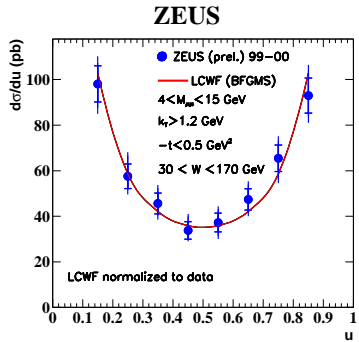}
%\vglue -1.0cm
\caption{Differential cross section $d\sigma/du$ measured for the selected
kinematic region (see text). The inner error bars show the statistical
uncertainty; the
outer error bars show the statistical and systematics added in quadrature.
The data points are compared to the prediction of LCWF theory. 
The theory is normalized to data.}
\label{res}
\end{center}
\end{figure}

This is the well known Bethe-Heitler process. Using the LCWF formalism it
has been shown \cite{ppnp} that the cross section for real photons and
using $m_l = 0 \rightarrow a = 0$ is:
                                                                                
\beq
\frac{d\sigma}{dt ~du ~dk_t^2} \propto
\frac{2[u^2 + (1-u)^2]}{k_t^4} \sim \frac{\Phi^2}{k_t^2}.
\eeq
This proportionality can be utilized in the same way as was done for the
pion. Measurement of the photon light-cone wave function was carried
out at the DESY accelerator in the collision of
28 GeV/c electrons (or positrons) with 920 GeV/c protons producing
real or virtual photons. The measurements were done with the ZEUS
detector \cite{zeus} using the exclusive $ep \rightarrow e  \mu^+ \mu^-
p$ photoproduction process. The integrated luminosity for the results
was 55.4$\,\pm\,$1.3 ${\rm pb^{-1}}$. The kinematic region was defined by
the following selection criteria:
the invariant mass of the dimuon system $4\,<\,M_{\mu\mu}\,<\,15\,$GeV
(above the resonances), the $\gamma p$ centre of mass energy
$30\,<\,W\,<\,170\,$GeV (region of stable and high acceptance), the square
of the four momentum exchanged at the proton vertex $|t|\,<\,0.5\,{\rm
GeV}^2$ (select diffractive events), $0.1\,<\,u\,<0.9$  (avoid the
end-points region with low acceptance) and $k_T\,>\,1.2\,$GeV (select a
hard process). The measured differential cross
section $d\sigma/du$ is presented in Fig. \ref{res} and is in good
agreement with the LCWF squared (BFGMS) \cite{bfgms}.
This measurement serves as ``Standard
Candle" and normalization for the Hadronic LCWF. It also provides the
first proof that diffractive dissociation of particles
can be reliably used to measure their LCWF. Furthermore it gives support
for the method used in previous measurements of the pion 
\linebreak LCWF \cite{791wf}
and possible future applications to other hadrons \cite{ppnp,com}.
\subsection{Measurement of the hadronic component of the photon LCWF
by Exclusive Dipion Electroproduction}
\label{pho_dpi}
\begin{figure}[ht]
\begin{center}
\includegraphics[width=6.0cm]{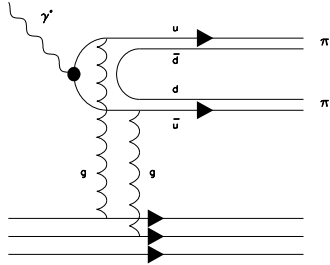}
\vglue -2.0cm
\caption{The photon dissociation to \qb followed by hadronization to
two pions}
\label{diag2pi}
\end{center}
\end{figure}

The processes of photo- or electro-production of two pions may be
considered as a special case of the photon dissociation to dijets when
each jet consists of one pion (Fig. \ref{diag2pi}). This is a very
exclusive process where the pion form factor and quantum numbers
may affect the ratio of longitudinal/transverse cross sections and the
$u$-distribution. The cross section for this process is expected to be
proportional to the time-like form factor:
\beq
\frac{\sigma(\gamma^*+p \rightarrow 2\pi + p)}{\sigma(\gamma^*+p
\rightarrow X + p)} \propto \left |F_{\pi} \right |^2
\eeq
where the denominator can be taken from parametrization of measurements
and the results may have to be normalized to those obtained from $e^+e^-$
measurements \cite{bisello}. It may be possible to use these measurements
to extend measurements of the pion time-like form factor into the
$4 < Q^2 < 15$ region where there is great sensitivity to the pion
light-cone wave function \cite{gous}.
The exclusive dipion reaction is considered as one of the ways to hunt for
the elusive Odderon \cite{hag,ginz} described as a d-coupled three-gluon
color singlet C = -1 object \cite{basa}. Since the charge parity of
dipions is: $C(\pi^+\pi^-) = (-)^{\ell}$ where $\ell$ is their relative
angular momentum, both Pomeron and Odderon can contribute to their
production and interference will show up. The signature would be
charge asymmetry: $A = \frac{u(+) - u(-)}{u(+) + u(-)}$. It  was
calculated for dipion electroproduction \cite{hag} and photoproduction
\cite{ginz}.
Measurement of exclusive diffractive electroproduction of $\pi^+\pi^-$
pairs was carried out by the ZEUS collaboration with integrated luminosity
of 66.3$\,\pm\,$1.7 ${\rm pb^{-1}}$. The kinematic conditions for this
measurement were: $2 < Q^2 < 20 GeV^2$,
$1.2\,<\,M_{\pi\pi}\,<\,5\,$GeV, $40\,<\,W\,<\,120\,$GeV,
$|t|\,<\,0.5\,{\rm GeV}^2$, $0.1\,<\,u\,<0.9$. The mass distribution, 
when divided by the mass dependence
of the $\gamma^* p \rightarrow q\bar{q}$ cross section can be used to
extend the measurements of the time-like form factor from the highest
available value at 2 Gev \cite{bisello} up to about 4 GeV.
The $t$ distribution is fitted with the
standard exponential form $d\sigma/dt \propto e^{bt}$. The resulting
slope $b= -4.81 \pm 0.42(stat)^{+0.05}_{-0.44}(sys) ~GeV^{-2}$ is in
good agreement with the values observed for vector-meson production
\cite{zeusvm}. This slope is considered as a measure of the combined size
of the \qb system and the
proton: $b = \frac{1}{3}<R^2> + \frac{1}{8}r^2$ where $r \sim
\frac{6}{\sqrt{Q^2 + m^2_V}}$ is the size of the \qb system before it
hadronizes to a vector meson or, in this case, to two pions in the
continuum and $R$ is the proton radius \cite{nik96}. The results
show that the \qb size in hadronization to continuum states is similar to
that obtained for vector mesons.
                                                                                
\begin{figure}[h]%\centering
\includegraphics[angle=0,width=7cm]{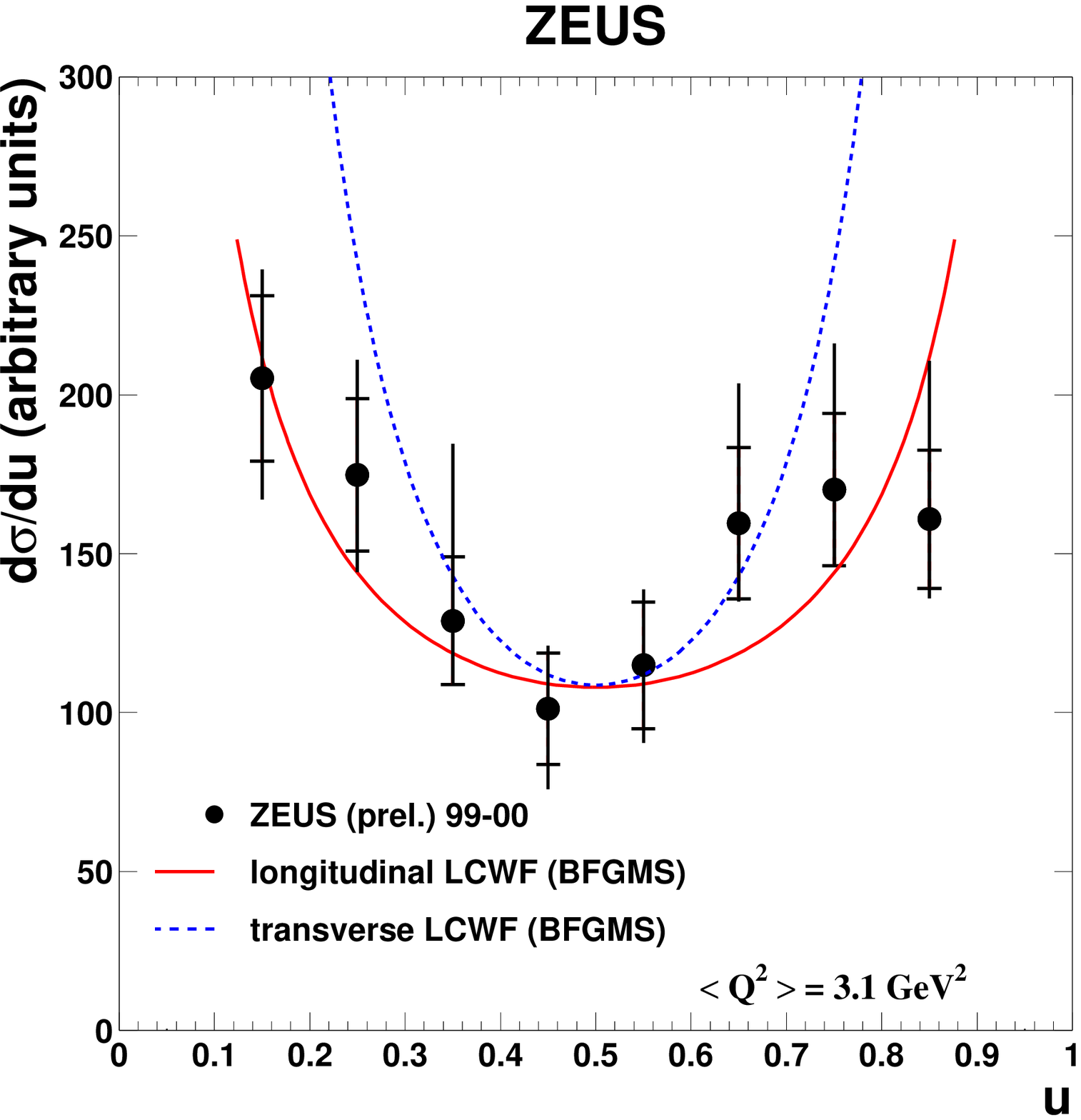}
\hfill
\hspace*{-2.0cm}
\includegraphics[angle=0,width=7cm]{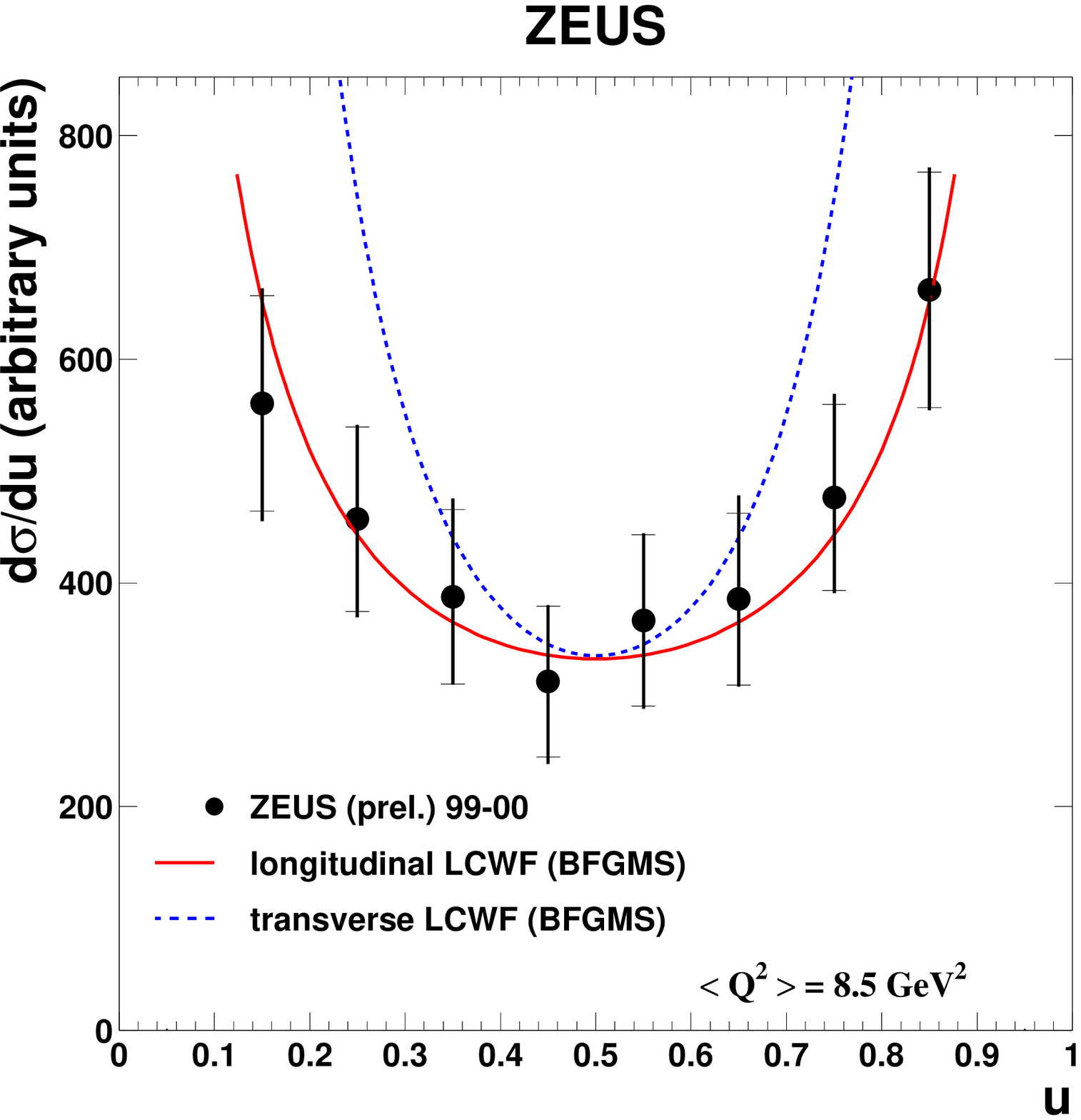}
                                                                                
\vglue -0.5cm
                                                                                
\caption{The differential cross section $d\sigma/du$ measured in two \q2
intervals: $2 < Q^2 < 5 ~GeV^2$ (left) and $5 < Q^2 < 20 ~GeV^2$ (right).
The inner error-bars show the statistical uncertainties; the outer
error-bars show the statistical and systematic uncertainties added in
quadrature. The data points are compared to the LCWF predictions.}
\label{upipi}
\end{figure}

The $u$ distribution is presented in Fig. \ref{upipi} for two \q2
intervals. Its shape is compared to the LCWF predictions for transverse
and longitudinal photons \cite{ppnp}. Such a comparison
is legitimate if the pion
quantum numbers do not affect their angular ($u$) distributions for a
given photon polarization. The normalization for the longitudinal LCWF
prediction was determined from fit to the data. For the transverse LCWF
prediction it was fixed to be the same value as that for the longitudinal
prediction at $u$ = 0.5. We note that the measured
distribution in the low \q2 region (Fig. \ref{upipi} left) is more
irregular than the one for the higher \q2 range. This irregularity can be
traced to the fact that the low \q2 range is close to the average
value of the dipion mass squared. In fact, for this rannge $<\beta> ~=
~0.52$ which is where $\sigma_L/\sigma_T ~\sim ~0$ in leading order
\cite{ppnp}. Higher order effects are expected to play a role here
and the pure longitudinal fluctuations may be shadowed. By contrast, for
the high \q2 range $<\beta> ~= ~0.75$ which is a ``safe" region.

The results are consistent with
the LCWF predictions for longitudinaly polarized photons. The
agreement between the measured $u$ distributions and the predictions
lends support to the assumption that non-resonant di-pion production is
sensitive to the \qb component of the light-cone wave function of the
virtual photon. This shows that for the phase space parameters of this
study the LCWF predicted by perturbative QCD
is correct. This process of exclusive diffractive electroproduction
of pion pairs can be correctly described as resulting from a
longitudinal photon fluctuating to a \qb pair which in turn hadronizes
to a $\pi^+ \pi^-$ pair, Fig. \ref{diag2pi}. The agreement of the
$u$-distribution
measured with pions in the continuum and in the resonance region above
the $\rho$ with calculations made at the parton level lend support to the
picture that there is parton/hadron duality, \cite{close}.

\end{document}